\documentclass[prb,twocolumn,showpacs]{revtex4-1}
\usepackage{amssymb,amsmath,bm,graphicx}

\begin{document}

\title{Snake states and their symmetries in graphene}
\author{Yang Liu$^2$, Rakesh P. Tiwari$^3$, Matej Brada$^2$,
  C. Bruder$^3$, F.V. Kusmartsev$^2$, and E.J. Mele$^{1,2} $}
    \email{mele@physics.upenn.edu}
    \affiliation{$^1$Department of Physics and Astronomy, University of Pennsylvania, Philadelphia PA 19104  \\
    $^2$Department of Physics, Loughborough University LE11 3TU, UK\\
    $^3$Department of Physics, University of Basel,
    Klingelbergstrasse 82, CH-4056 Basel, Switzerland}
\date{\today}

\begin{abstract}
  Snake states are open trajectories for charged particles propagating
  in two dimensions under the influence of a spatially varying
  perpendicular magnetic field. In the quantum limit they are
  protected edge modes that separate topologically inequivalent ground
  states and can also occur when the particle density rather than the
  field is made nonuniform. We examine the correspondence of snake
  trajectories in single-layer graphene in the quantum limit for two
  families of domain walls: (a) a uniform doped carrier density in an
  antisymmetric field profile and (b) antisymmetric carrier
  distribution in a uniform field. These families support different
  internal symmetries but the same pattern of boundary and interface
  currents. We demonstrate that these physically different situations
  are gauge equivalent when rewritten in a Nambu doubled formulation
  of the two limiting problems. Using gauge transformations in
  particle-hole space to connect these problems, we map the protected
  interfacial modes to the Bogoliubov quasiparticles of an interfacial
  one-dimensional p-wave paired state. A variational model is
  introduced to interpret the interfacial solutions of both domain
  wall problems.
\end{abstract}

\maketitle

\section*{Introduction}
A charged particle moving in two dimensions under the influence of a
spatially varying perpendicular magnetic field can exhibit snake state
trajectories. These are open two-dimensional orbits perpendicular to
the direction of the magnetic field gradient. Snake trajectories occur
in both the classical and quantum limits of this problem and are of
fundamental interest with potential applications for electron
transport in multidomain ferromagnets, two-dimensional electron
gases~\cite{muller,peeters}, and in nanomaterials like
graphene~\cite{ghosh2008,park2008,oroszlany2008}.  In the
quantum limit the snake states can be interpreted as the protected
modes that occur at domain walls that separate topologically mismatched
gapped ground states. This picture suggests that snake trajectories
can arise even in a {\it uniform} magnetic field if the particle
density is suitably modulated laterally, e.g. by electrostatic gating
patterned to form interfaces between distinct quantum Hall ground
states.

Indeed exactly this possibility has been explored
theoretically~\cite{breyfertig1,williams,abaninssc,abaninscience} and
examined experimentally for graphene in a uniform perpendicular
magnetic field via measurements of the Hall conductance and of
Fabry-Perot like oscillations in the inter-edge conductance across
graphene {\it pn} junctions~\cite{schonenberger,ozylimaz}.  Graphene
is an excellent candidate for this application because it can be
electrostatically switched from {\it n} to {\it p} carrier types and
studied in the ballistic transport regime~\cite{b1}. The converse
problem of snake trajectories for a uniform carrier density in a
spatially varying magnetic field is even more technically challenging
and it has not been examined experimentally (for a theoretical
discussion see Ref.~\onlinecite{prada}). On the other hand, a variant
of this latter problem {\it is} routinely encountered in present-day
experimental environments. In single-layer graphene subject to elastic
lattice strains, the low-energy electronic structure is described by a
Dirac Hamiltonian containing a strain-induced gauge field that mimics
the effects of a perpendicular (albeit valley antisymmetric) magnetic
field.  For generic smoothly varying strain fields the presence of
nodal lines that separate regions of ``positive" and ``negative"
pseudo-magnetic field in a single valley is a nearly unavoidable
consequence of the symmetry of this strain coupling.

The ``antisymmetric $B$" and ``antisymmetric doping ($V$)" problems
break both time reversal symmetry ${\cal T}$ (due to the presence of a
magnetic field) and particle-hole symmetry $\Xi$ (they require a
nonzero carrier density). Nonetheless they retain different composite
symmetry operations that combine these discrete symmetries with
twofold rotation about the layer normal ${\cal R}_z$: the former
problem is symmetric under ${\cal R}_z {\cal T}$ and the latter under
${\cal R}_z \Xi$. This difference manifests itself in many of the
spectral properties presented in Section I. In fact, this distinction
persists even into the classical limit, and reflects the different
underlying dynamics of these two problems. In the antisymmetric-$B$
problem a snake state trajectory arises from the compensation of the
circulation of cyclotron orbits in regions where the magnetic field is
reversed. In the antisymmetric-$V$ it arises from one-sided skipping
orbits due to an electric field at the interface of a $pn$ junction.
It is perhaps surprising that these problems exhibit the {\it same}
pattern of boundary and interface currents. This can be understood as
a consequence of confinement of these boundary modes at the interface
between topologically mismatched gapped ground states on either side
of the interface.  It is therefore of interest to understand precisely
how these different problems are related in the bulk. In this paper we
observe that these two situations are in fact gauge equivalent
representations of the same problem. However our demonstration of this
equivalence requires that we extend both problems in a Nambu-doubled
formulation, explicitly restoring particle hole symmetry about a
nonzero chemical potential.  In this doubled representation we find
that the problems are interconverted by local gauge transformations
exploiting the particle and hole degrees of freedom in the Nambu
basis. Among the insights provided by this approach, we observe that
the interfacial degrees of freedom common to the two problems (the
snake states) are mapped to a model for the Bogoliubov quasiparticles
in a one-dimensional superconductor along the tangent line that
supports a ``p-wave" pairing field.

This paper is organized as follows. In Section I we begin by comparing
the spectra for antisymmetrically doped graphene (a {\it pn} junction) in
uniform field with the spectrum for uniform doping in an antisymmetric
field. In Section II we present a family of local gauge
transformations that allow one to map one problem onto the other. In
Section III we implement this procedure for the case of the graphene
{\it pn} junction and analyze the structure of the uniformly doped
antisymmetric field problem to which it maps. In Section IV we present a
topological analysis of the ground state manifolds in these models,
concluding that they are the same and are members of the
Altland-Zirnbauer chiral symmetry class C, i.e. they are indexed by
even integer-valued winding numbers. In Section V we use a variational
approach to determine the spatial structure of the interface modes.

\section{Spectra of the folded graphene and pn junction in magnetic field}

We consider snake state solutions in two limits of a tight-binding
theory for electrons on a honeycomb lattice.  The Hamiltonian is
\begin{eqnarray}
{\cal H} = \sum_i \, (V_i - \mu) \left(c^\dag_i c_i-1/2 \right) + \sum_{\langle i,j \rangle} \, t e^{i \varphi_{ij}} c^\dag_i c_j + {\rm h.c.} \nonumber\\
\label{tb_hamiltonian}
\end{eqnarray}
where $\varphi_{i,j} = (e/\hbar) \int_i^j \vec A \cdot d \vec \ell$ is
the Peierls phase accumulated in a (possibly nonconstant)
perpendicular magnetic field $B_i$. We adopt a coordinate system where
the scalar potential $V_i$ and the vector potential $\vec A_i$ are
spatially varying in the $x$ direction and constant along $y$.  The
chemical potential $\mu$ is set so that the left and right sectors are
simultaneously gapped. The calculations are carried out for ``zigzag"
interfaces, where a domain wall at $x=0$ is tangent to a primitive
translation vector along $y$. We examine two limiting domain wall
geometries. In the first we assume that the system is uniformly doped,
say $p$-type on both sides of an interface where $B(x)$ changes
sign. In the second we consider the complementary case where the $B$
field is uniform (or at least a symmetric function of $x$ with no sign
changes), and instead the external scalar potential $V(x)$ with zero
mean changes its sign on an interface defining a $pn$ junction.  
We show that the edge-state solutions for these two limits are the same
despite the different microscopic dynamics. This manifests a
topological equivalence of their bulk ground states.  Indeed we find
that these can be mapped into each other by gauge transformations that
mix the particle and hole degrees of freedom when the problem is
rewritten in a Nambu particle-hole basis.  This leads to the
possibility of inventing architectures that simulate 
unusual ballistic transport effects like Andreev reflection even in
the absence of a physical superconducting condensate. In the
following, we will first discuss the two limits separately and then
analyze their gauge equivalence.

\subsection{Antisymmetric B, Symmetric V}
\begin{figure}
  \includegraphics[angle=0,width=\columnwidth]{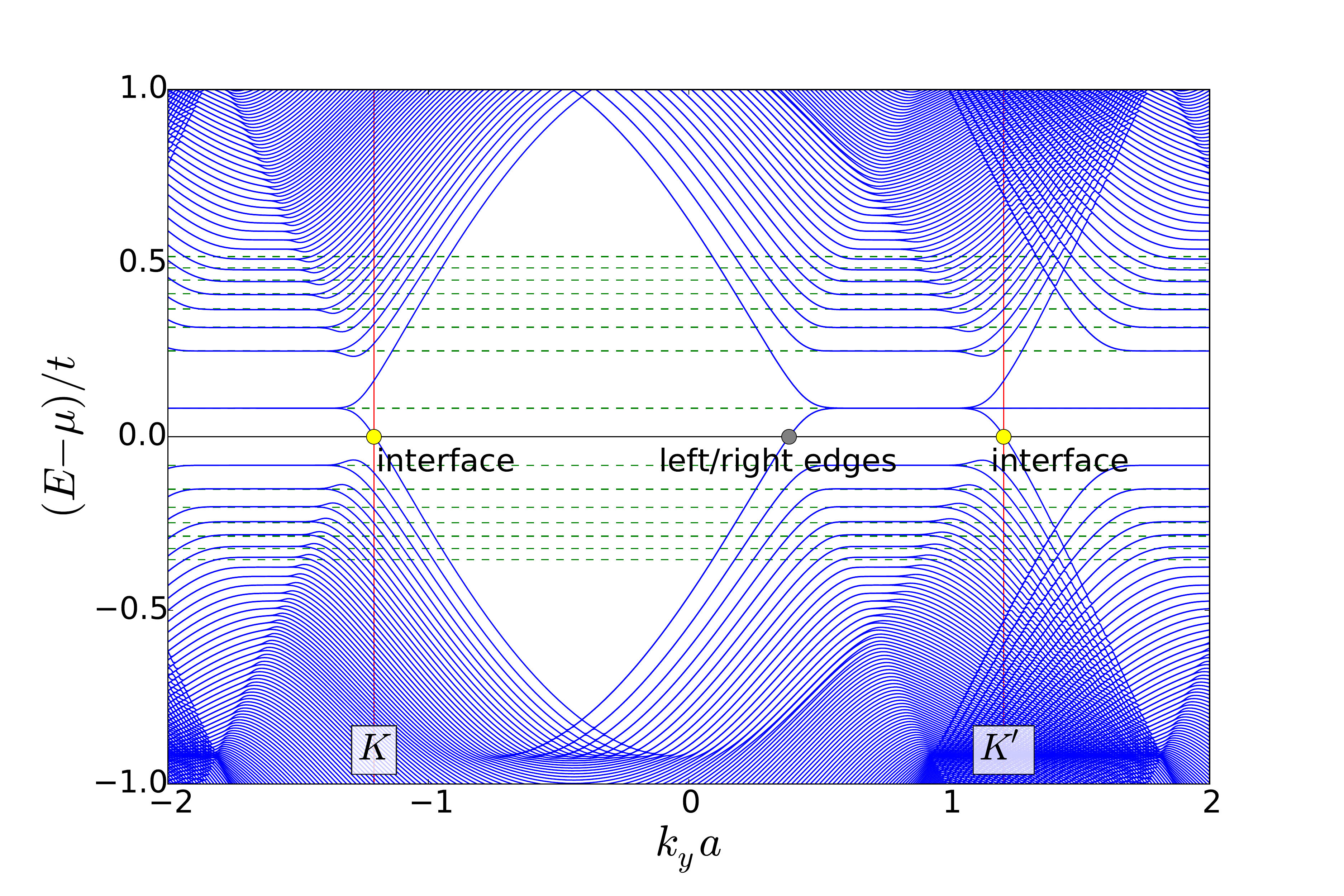}
  \caption{\label{fig1} Numerically calculated spectrum for graphene
    with uniform doping in an antisymmetric magnetic field profile. We
    plot $E-\mu$ as a function of $k_ya$, where $a$ is the
    inter-atomic distance in graphene, and assume $p$-type doping
    (i.e., $\mu=0$, and a uniform scalar potential $V$ that shifts the
    system away from neutrality).  The spectrum shows four dispersing
    features at the Fermi energy: two with ``positive" velocity on the
    outer edges, and a pair of modes with ``negative" velocity in the
    domain wall. The flat band that extends from the $K$ to $K'$
    points is the surface state for a zigzag edge. The horizontal
    dashed lines show the dispersionless bulk Landau levels for a
    Dirac system.}
\end{figure}

We assume that the system is uniformly slightly doped $p$-type and has
an antisymmetric magnetic field profile $B(x) = B_o \tanh (x/\ell)$.
The spectrum for this problem is displayed in Fig.~\ref{fig1}. The
vertical red lines denote the projections of the bulk $K$ and $K'$
Dirac points, i.e. in the absence of a field these are the
interface-projected locations of the bulk gap closures.

For an antisymmetric $B(x)$ the vector potential in Landau gauge is an
even function with $A_y(-x) = A_y(x)$. Consequently the system
supports normalizable (near) zero-energy states that are ``one-sided"
in momentum space as shown in Fig.~\ref{fig1}.  There are two types
of momentum-space anisotropy evident in these spectra: (a) The
$B$-induced zero modes occur only for $q = k_y-K(K') <0$ in both
valleys (b) The $q<0$ spectrum near $K'$ supports an additional pair
of zero modes due to the undercoordinated atoms at the zigzag edges
(the total orbital degeneracy of the $q<0$ spectrum is actually four
in this region). The additional low-energy modes bridge the $K'$ and
$K$ points where they smoothly evolve into the field-induced zero
modes in the opposite valley. In either case the transition from $q<0$
to $q>0$ marks a crossover where the zero-energy degrees of
freedom hybridize to produce a pair of particle-hole symmetric
propagating modes that are confined to the domain wall.  For
$p$ type doping (as illustrated) this pair of interfacial modes
copropagate along $- \hat y$.  Physically the pair of domain wall
modes combine cyclotron orbital states of opposite circulation to confine
their motion near the interface.

\begin{figure}
  \includegraphics[angle=0,width=\columnwidth]{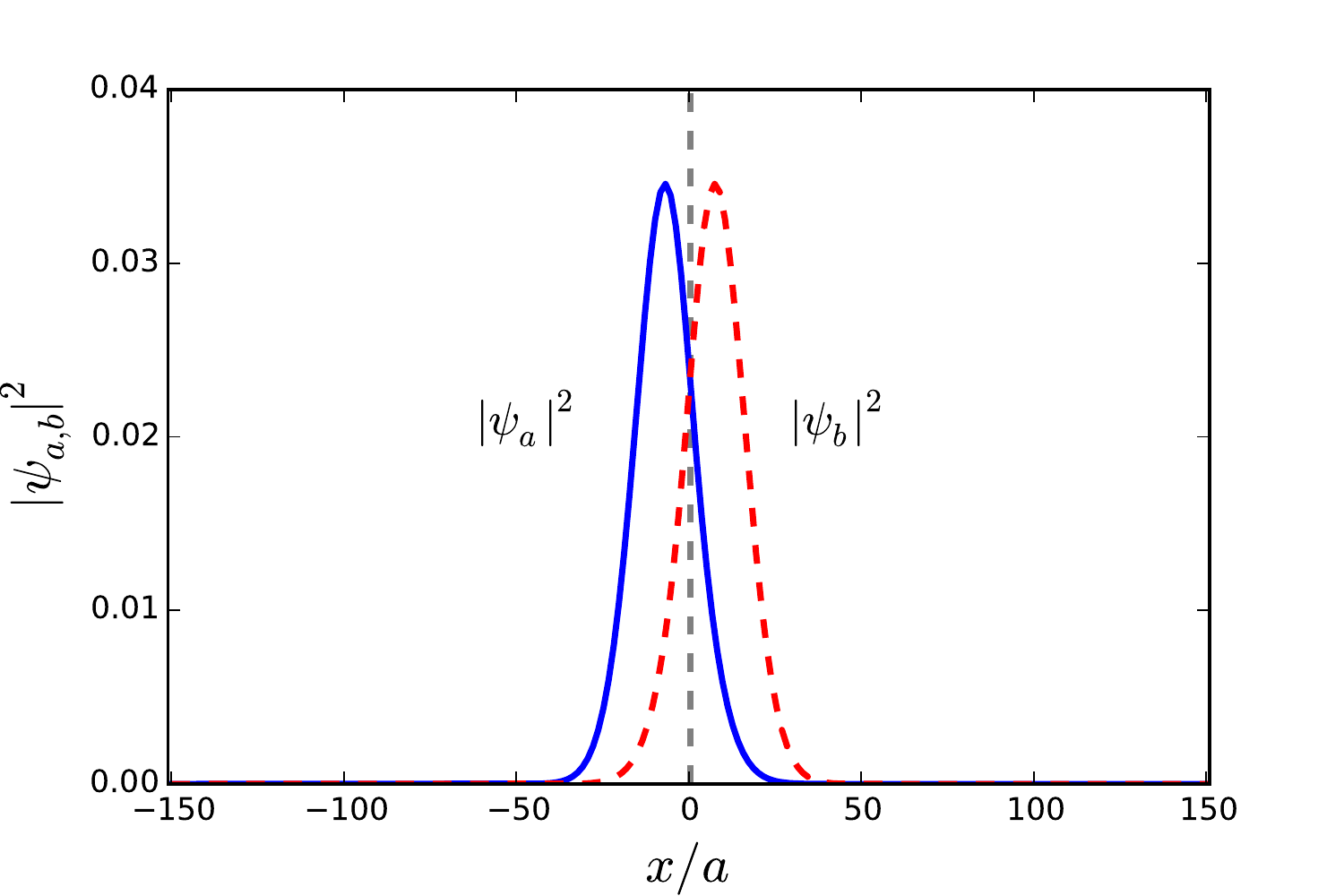}
  \caption{\label{fig2} Numerically calculated probability (or charge)
    density for one of the interface zero-energy modes shown in
    Fig.~\ref{fig1}. The mode is associated with valley $K$ for
    uniform doping in an antisymmetric magnetic field profile, and
    $\psi_a$ (solid blue) and $\psi_b$ (red dashed) are the two spinor
    components.  For the other valley, $K'$ the density of the $a$ and
    $b$ components of the wave function are interchanged. The $x$-axis
    is measured in units of the interatomic distance $a$ in graphene.}
\end{figure}

\begin{figure}
  \includegraphics[angle=0,width=\columnwidth]{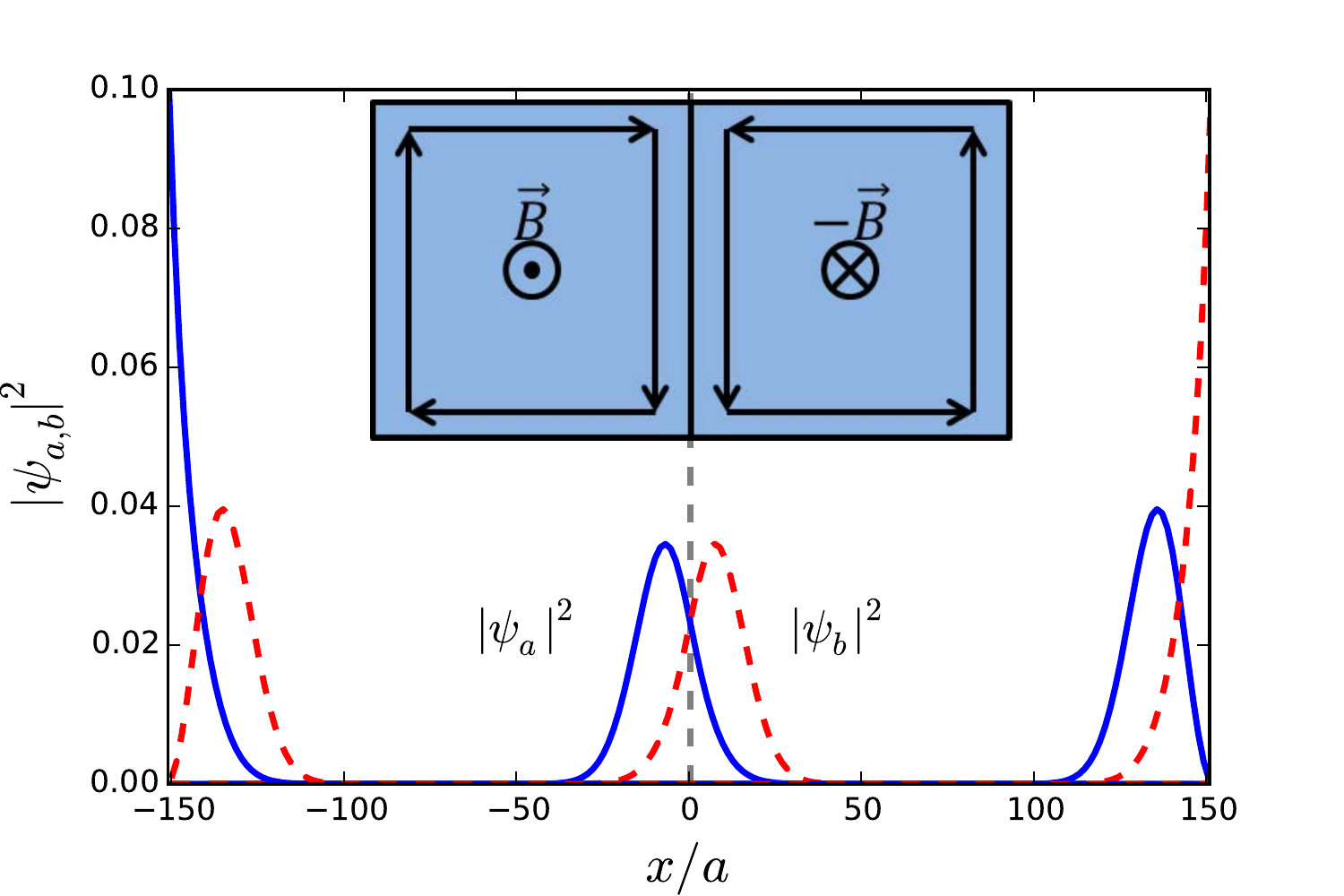}
  \caption{\label{fig3} Numerically calculated probability (or
    charge) density for the zero-energy modes shown
    in Fig.~\ref{fig1}. The two peaks in the center correspond to the
    interface modes of Fig.~\ref{fig2}. The peaks on the left
    and right side correspond to the edge modes. The solid blue (red
    dashed) curve represent the $\psi_a$ ($\psi_b$) spinor component.
    Inset: Schematic of the current pattern due to opposite cyclotron motions
    in the two halves of the setup.}
\end{figure}

The dispersion of the outer-edge modes near the right-hand $K'$ valley
is particularly instructive. Note that this band is nearly flat for
small $q<0$ but it becomes strongly dispersive with positive group
velocity for sufficiently large negative $q$.  This occurs via
hybridization of the sublattice-polarized edge degree of freedom with
the Landau zero mode {\it on the opposite sublattice} when their
guiding centers are forced to the outer edges of the ribbon. Note that
the antisymmetry of $B(x)$ requires that the guiding centers are
forced to opposite outer edges at the same value of the crystal
momentum $k_y$. These dispersive outer edge modes constitute a return
path for the topological current induced in the domain wall.  These
features can be identified in the spatial distribution of the
probability (or charge) densities plotted in Figs.~\ref{fig2}
and \ref{fig3}.

To summarize, for constant $V$ and antisymmetric $B$ we observe
(a) Four interface/edge modes at the Fermi energy.
(b) A pair of co-propagating modes at the domain wall which
    combine cyclotron motions of opposite circulation.
(c) Outer edge modes that hybridize the zeroth Landau level with
    the zigzag surface state.
(d) Valley asymmetry: domain wall modes occur in both valleys but
    there is support for the outer edge modes only in a single
    valley. Reversal of the direction of $B$ everywhere will select
    the other valley.

\subsection{Symmetric B, Antisymmetric V}

We now consider the opposite limit that occurs with uniform magnetic
field and an antisymmetric bias $V(x) = V_o \tanh(x/\ell)$. This
creates a graphene $pn$ junction in a uniform field which is the
situation studied in two recent
experiments~\cite{schonenberger,ozylimaz}.

\begin{figure}
  \includegraphics[angle=0,width=\columnwidth]{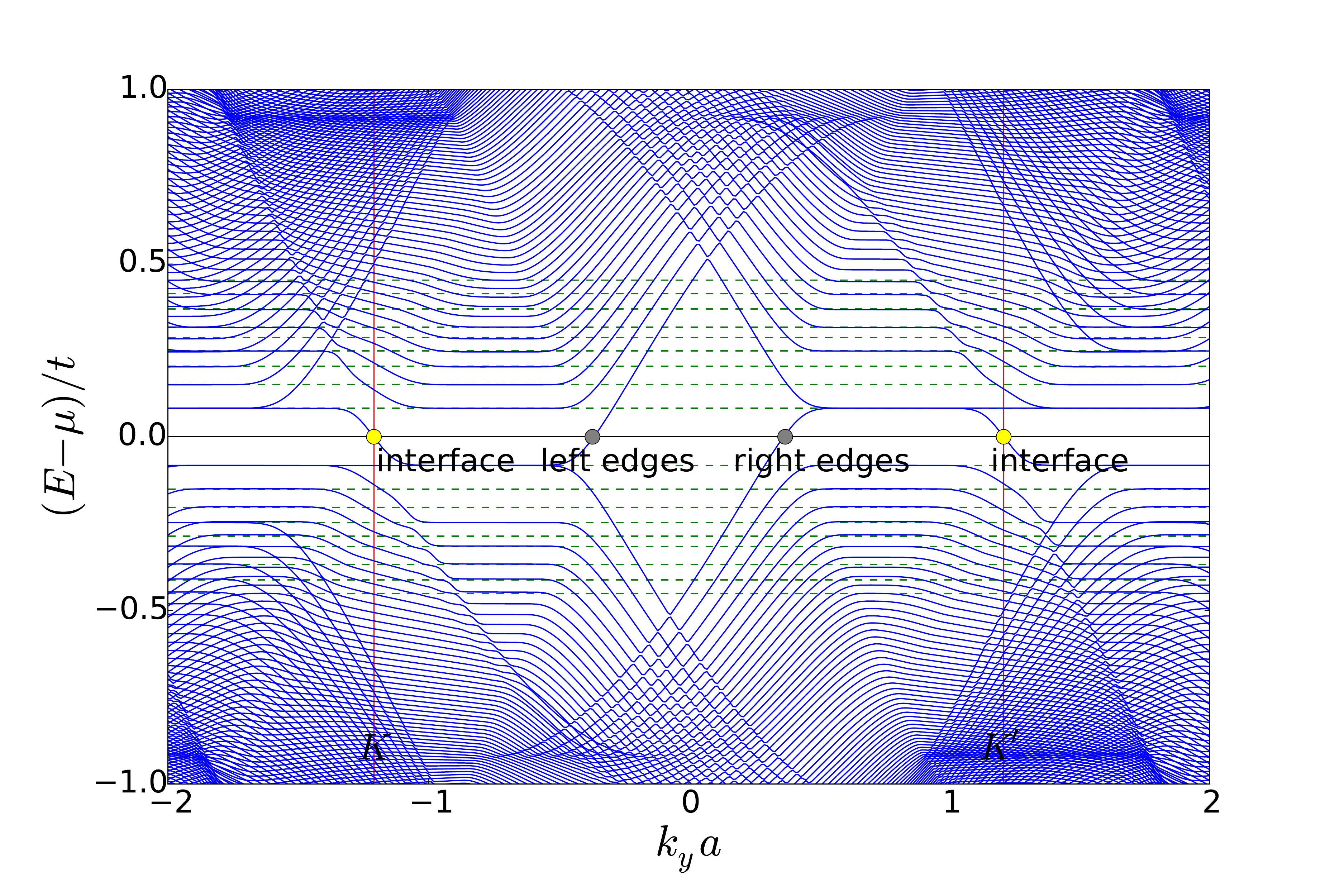}
  \caption{\label{fig4} Spectra for the $pn$ junction in a uniform
    field. The plot gives $E-\mu$ as a function of $k_y$ for a
    geometry where $B$ is constant but $V(x) = V_o \tanh(x/\ell)$. The
    system supports four edge and interface modes: two positive
    velocity modes on the outer edges and two negative velocity modes
    at the domain wall. The flat feature is a zigzag edge state that
    morphs into the zeroth Landau level. The spectra are one-sided,
    and show graphene character near a shifted neutrality point of one
    sign of $q=k_y-K$ in one valley and the opposite sign in the other.
    At $k_ya=\pi/\sqrt{3}$ the spectra are twofold degenerate.}
\end{figure}

The spectrum calculated for this configuration is displayed in
Fig.~\ref{fig4} where we plot $E-\mu$ as a function of $k_y$. Here
the system is $n$-doped for $x<0$ and $p$-doped for $x>0$. Again one
finds four dispersing modes at the Fermi energy: two with negative
velocity at the domain wall, and two with positive velocity confined
on the outer edges.  Despite this similarity, the mechanism producing
the edge state structure is quite different.  We note that the
$\sqrt{n}$ signature of the Landau quantization of the Dirac spectrum is
observed for $q=k_y-K>0$ in the $K$ valley but for $q'=k_y-K'<0$ in
the $K'$ valley, i.e. the Landau quantized spectra are both one-sided
in momentum space, but with opposite senses in the two valleys. In the
forbidden regions $q<0$ and $q'>0$ the spectrum collapses to a pair of
nearly degenerate orbital doublets that connect the two valleys. This
degeneracy is exact at $k_ya=\pi/\sqrt{3}$: the energy jump that is produced
by the transition $q>0$ to $q<0$ (and vice versa for $q'$) is the
quantized energy spacing between the zeroth and first Landau levels,
so all the levels for $k_ya =\pm \pi/\sqrt{3}$ are twofold degenerate.

Here the dispersion of the confined interfacial modes can be
understood as a response to the lateral electric field produced in the
$pn$ junction. As $x$ crosses zero the scalar potential $V(x)$
switches its sign and the internal electric field $E=-\partial_x V$ is
nonzero. Thus a state with drift velocity $\vec E \times \vec B/B^2$
sees no deflection and can propagate freely. This can be contrasted
with the guiding-center mechanism that liberates these modes in the
former antisymmetric-$B$ problem where $E=0$ and one requires the
compensation of the circulation in orbits in reversed $B$ fields to
produce freely propagating interfacial snake states.

\begin{figure}
  \includegraphics[angle=0,width=\columnwidth]{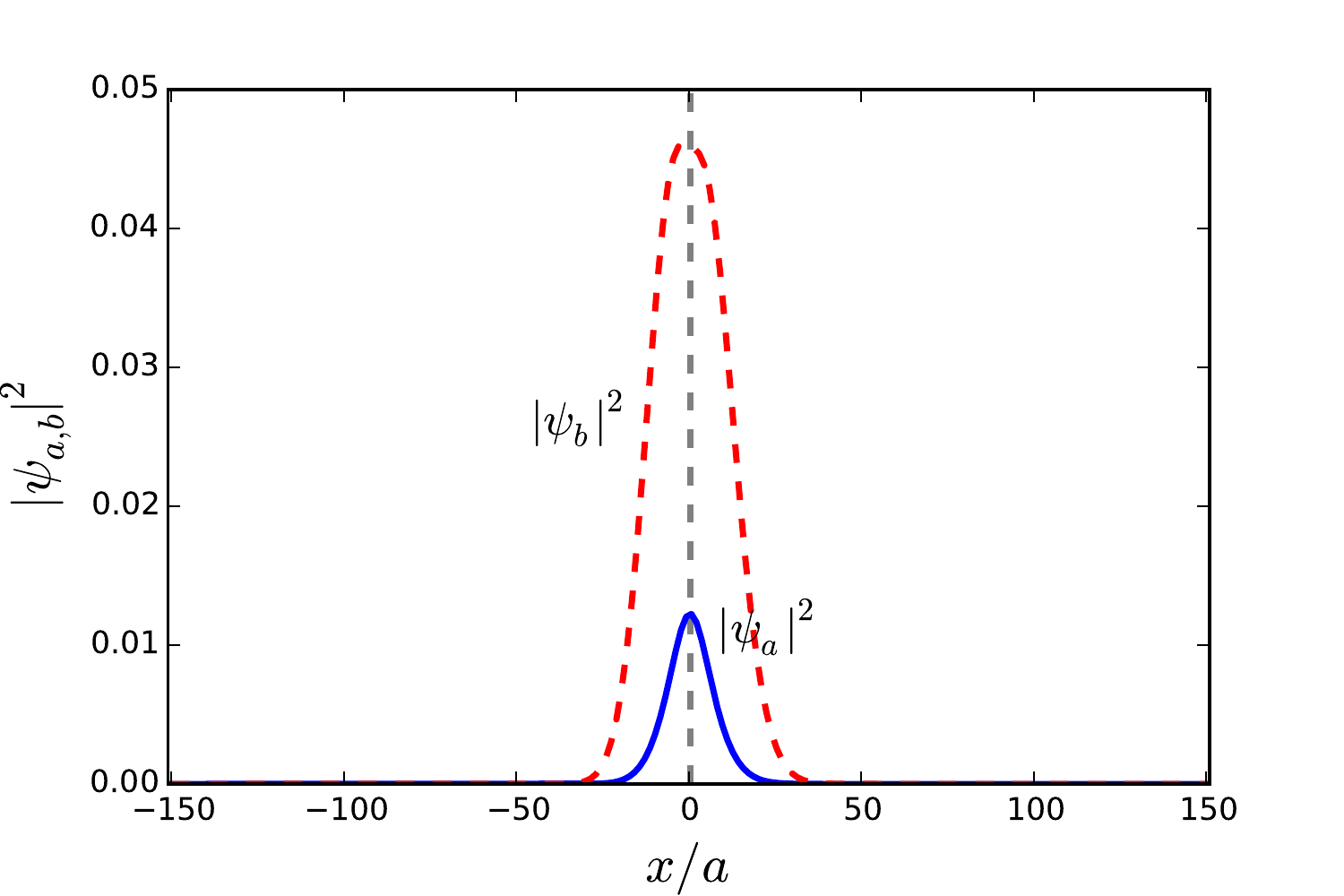}
  \caption{\label{fig5} Numerically calculated probability (or charge)
    density for one of the interface zero-energy modes shown in
    Fig.~\ref{fig4}. The mode is associated with valley $K$ for a
    graphene {\it pn} junction with the potential profile
    $V(x) = V_o \tanh(x/\ell)$ in a homogeneous perpendicular magnetic
    field, $\psi_a$ and $\psi_b$ are the two spinor components.  For
    the other valley, $K'$ the density of the $a$ and $b$ components
    of the wave function are interchanged.}
\end{figure}

\begin{figure}
  \includegraphics[angle=0,width=\columnwidth]{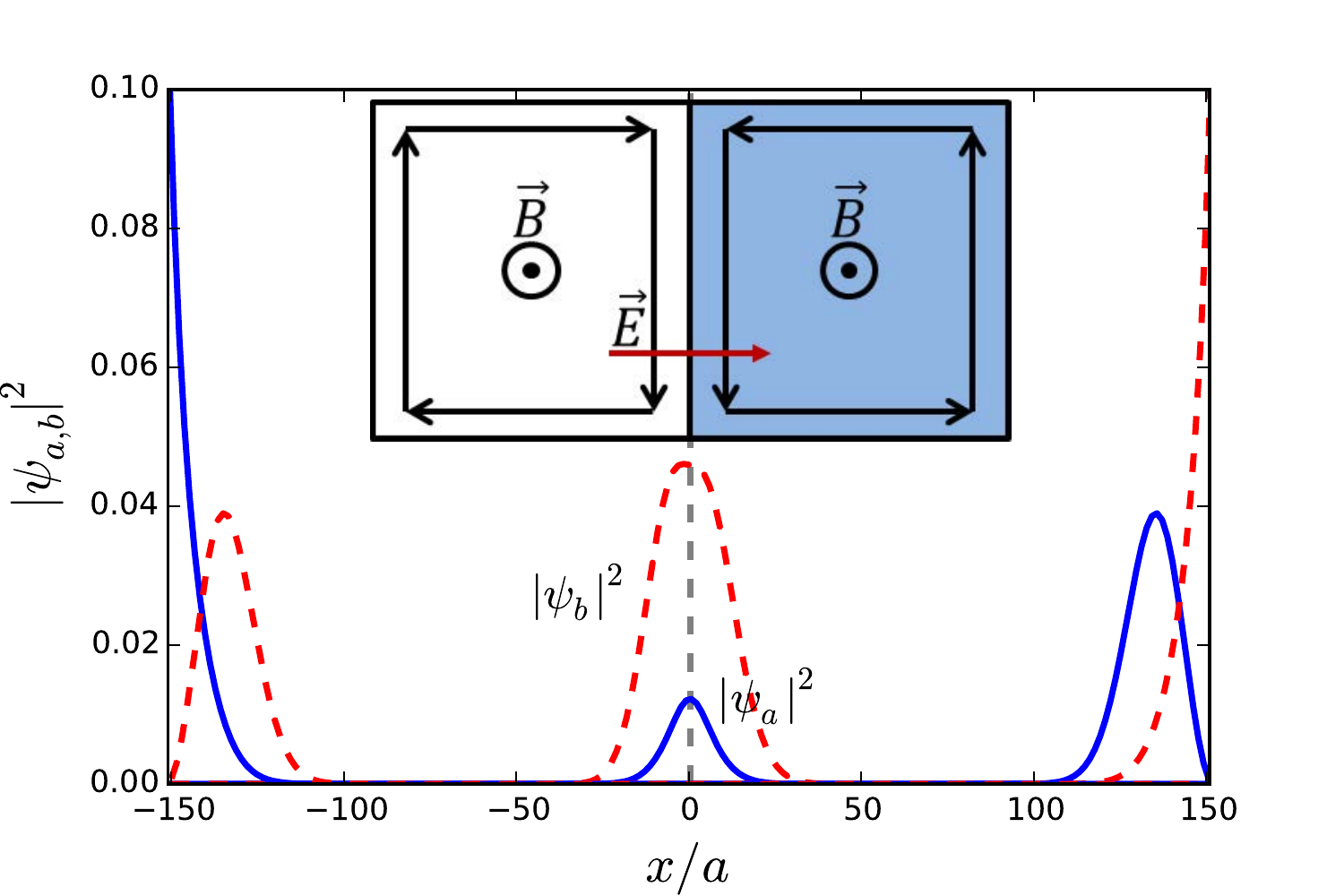}
  \caption{\label{fig6} Numerically calculated probability (or charge)
    density for the zero-energy modes shown in Fig.~\ref{fig4}. The
    two peaks in the center correspond to the interface modes of
    Fig.~\ref{fig5}. The peaks on the left and right side correspond
    to the edge modes. The solid blue (red dashed) curve represent the
    $\psi_a$ ($\psi_b$) spinor component.  Inset: Schematic of the
    current pattern showing the two co-propagating modes at the
    interface between the two half of the setup. The electric field
    $\vec E$ corresponds to the potential gradient at the $pn$ junction.}
\end{figure}

Interestingly, the appearance of dispersive edge modes on the {\it
  outer} boundaries follows exactly the same recipe as for the
antisymmetric $B$ problem. The guiding center of the Landau zero mode
which is sublattice polarized, is forced to the outer edge of the
ribbon where it hybridizes with the zigzag surface state on a
complementary sublattice to form the one-way dispersive
excitation. However, because the $B$ field is constant in this
problem, the guiding centers are forced to the outer edges of the
ribbon at opposite momenta $\pm k_y$. The entire spectrum of
Fig.~\ref{fig4} is then invariant under the combined transformation
$E-\mu \rightarrow -(E-\mu)$ and $k_y \rightarrow -k_y$.  The
probability (or charge) densities associated with the zero modes is
plotted in Figs.~\ref{fig5} and \ref{fig6}. We note that
the charge density shows a sublattice polarization, favoring the
sublattice found in Landau zero mode. This sublattice polarization is
captured in the variational approach presented in Section V.

To summarize the main results from this model for the graphene 
$pn$ junction in a uniform $B$:
(a) Four interface/edge modes at the Fermi energy.
(b) Co-propagating modes in the domain wall determined by their drift velocity
 specified by $B$ and the potential gradient in the wall.
(c) Conventional outer edge modes that hybridize a Landau zero
    mode with the surface state.

\section{Gauge equivalence via Nambu formulation}

It is striking that despite the different microscopic origins of the
domain wall solutions and the different structure of the full spectra
apparent in Figs.~\ref{fig1} and \ref{fig4}, the basic pattern of
the edge state currents is the same. This is evidence of the
topological character of these modes. By negating either $B$ or $V$ at
the interface we reverse the sign of the Chern number in the first
fundamental gap between Landau levels and therefore we require the
same pattern of boundary currents. This suggests that the ground
states of these two systems can be adiabatically mapped into each
other. This conclusion is surprising since the momentum-space
structures of their spectra examined in the previous section are
evidently controlled by the underlying dynamics which are quite
different and in fact incompatible for the two states.  In this
section we show that their ground states can nonetheless be mapped
into each other using a particle-hole extension of the original
formulations of both problems. The required mapping is a rotation in
the particle and hole degrees of freedom expressed in a Nambu
basis. Local gauge transformations in this basis interconvert the two
problems at the expense of introducing a fictitious pairing field
within the domain wall. In this section we develop a family of such
mappings and discuss the consequences of the induced
pseudo-pairing field in Section III.

The problems of Section I.A and I.B are distinguished by the coupling
of external potentials to bilinear terms in the fermion operators $c$ and
$c^\dag$. For example, the Peierls phase in Eq.~(\ref{tb_hamiltonian}) is
coupled to nearest-neighbor bilinear terms in the form
\begin{eqnarray}
t e^{i \varphi_{i,j}} c^\dag_i c_j\:,
\label{peierls}
\end{eqnarray}
while the scalar potential that defines the local doping is coupled
through the site density operator
\begin{eqnarray}
V_i \left(c^\dag_i c_i - 1/2 \right)\:.
\label{charge}
\end{eqnarray}
Sign reversal of the magnetic field direction negates the
exponentiated phase in Eq.~(\ref{peierls}) while a reversal of the
scalar potential flips the sign of the coupling to the net charge
operator Eq.~(\ref{charge}). Ignoring the physical spin of the
electrons, these reversals are introduced by the particle-hole
transformation
\begin{eqnarray}
\bar c = c^\dag \nonumber\\
\bar c^\dag = c
\end{eqnarray}
whereby
\begin{eqnarray}
t e^{i \varphi_{i,j}} c^\dag_i c_j \mapsto -t e^{i \varphi_{i,j}} \bar c^\dag_j  \bar c_i =  -t e^{-i \varphi_{j,i}} \bar c^\dag_j  \bar c_i \nonumber\\
V_i \left(c^\dag_i c_i - 1/2 \right)\mapsto -V_i \left(\bar c^\dag_i \bar c_i - 1/2 \right)\:.
\label{reversal}
\end{eqnarray}
For our application it is useful to collect these operators in
two-component spinors that resolve the two degrees of freedom at each
Bloch wavevector $k$
\begin{eqnarray}
\psi_k = \left(a_k, \, b_k\right)
\end{eqnarray}
and write the original problem in a doubled Nambu four-component basis
$\Psi_k = (\psi_k, \psi^\dag_{-k})$
\begin{eqnarray}
{\cal H}_N(k) = \left(
                    \begin{array}{cc}
                      \psi^\dag_k & \psi_{-k} \\
                    \end{array}
                  \right)
                  \left(
                    \begin{array}{cc}
                      {\cal H}_{k} & 0 \\
                      0 & -{\cal H}_{-k}^* \\
                    \end{array}
                  \right)
                  \left(
                    \begin{array}{c}
                      \psi_k \\
                      \psi^\dag_{-k} \\
                    \end{array}
                  \right) \:.
\label{H_Nambu}
\end{eqnarray}
When summed over $k$ this theory gives a doubled ``redundant"
description of the original problem.

Using Eqs.~(\ref{reversal}) a global rotation in particle-hole space
can globally ``choose" the signs of $B$ and $V$.  For example,
consider a Hamiltonian ${\cal H}[V(x),B(x)]$ parameterized by the
fields $V(x)$ and $B(x)$. Then define $2 \times 2$ Pauli matrices
$\sigma_\mu$ acting on the two sublattice degrees of freedom and
$\Sigma_\mu$ acting on the particle and hole degrees of freedom in the
Nambu representation.  A global operator of the form
\begin{eqnarray}
{\cal S}(\theta) = \cos \theta \, {\cal S}_1 + \sin \theta \,{\cal S}_2\:,
\label{S_theta}
\end{eqnarray}
where ${\cal S}_1 = \sigma_3 \otimes \Sigma_1$ and
${\cal S}_2 = \sigma_3 \otimes \Sigma_2$, has the property of formally
flipping the signs of $V$ and $B$ everywhere in the manner
\begin{eqnarray}
{\cal S}(\theta)^\dag {\cal H}[V(x),B(x)] {\cal S}(\theta)
  = {\cal H}[-V(x),-B(x)]\:.
\end{eqnarray}
We will discuss the behavior of ${\cal S}(\theta=0) = {\cal S}_1$
first and then return to the interpretation of the remaining phase
degree of freedom $\theta$.

\begin{figure}
  \includegraphics[angle=0,width=\columnwidth]{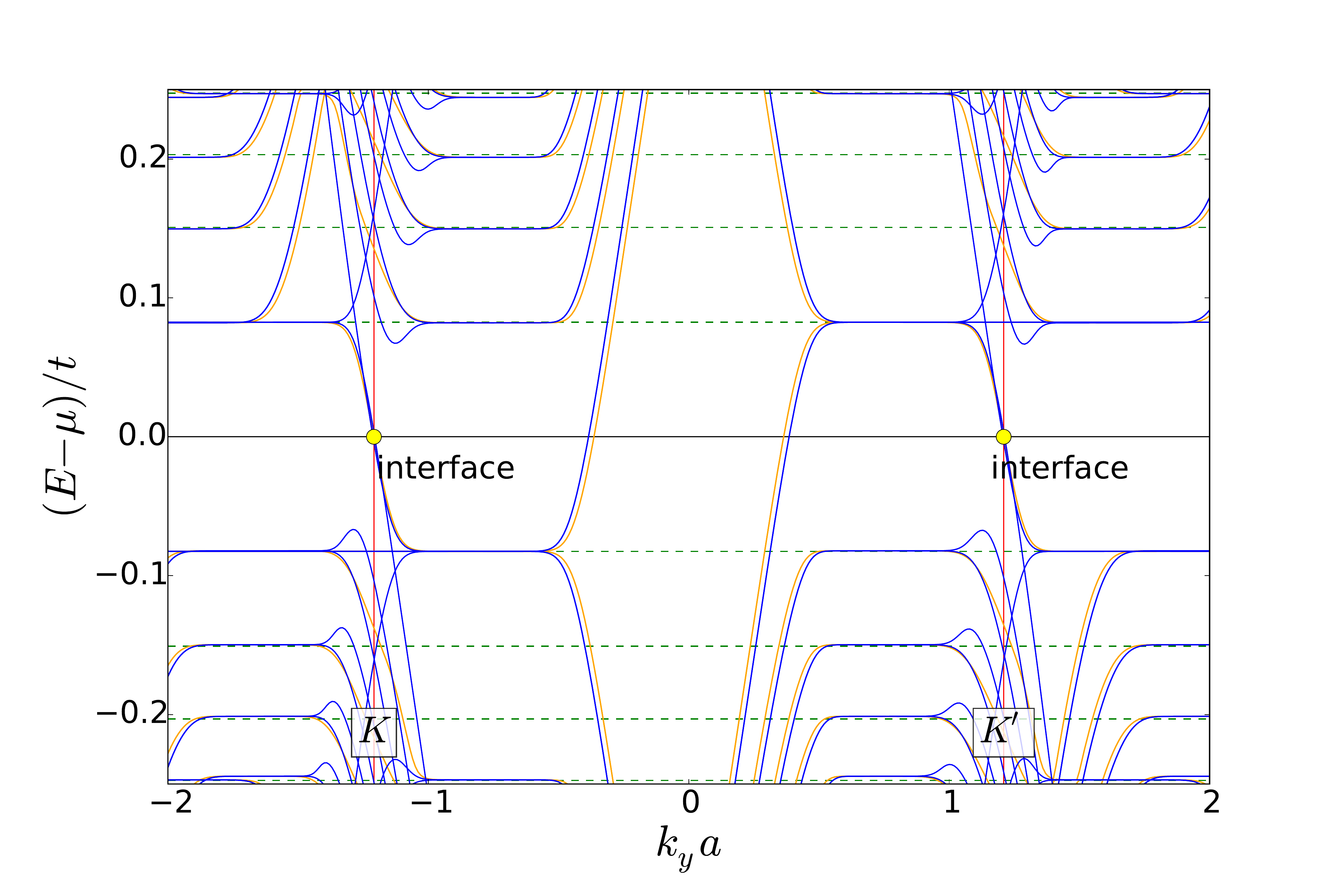}
  \caption{\label{fig7} Bogoliubov spectrum for the topological domain
    wall in graphene corresponding to both Figs.~\ref{fig1} and
    \ref{fig4}. The spectrum is the particle-hole doubled version of
    the antisymmetric $B$ spectrum, which is plotted only in the
    particle channel in Fig.~\ref{fig1}. The blue lines correspond to
    the antisymmetric magnetic field spectrum as shown in Fig.~\ref{fig1}; the
    orange lines correspond to the spectrum of a $pn$ junction in a
    uniform magnetic field as shown in
    Fig.~\ref{fig4}. The horizontal dashed green lines show the
    Landau-level spectrum for a Dirac particle in a uniform magnetic
    field for a chemical potential lying symmetrically between the
    zeroth and the first Landau level.}
\end{figure}

In an analogous manner, if we promote ${\cal S}$ to a local gauge
degree of freedom we can introduce a gauge transformation
${\cal S}_1(x)$ that {\it locally} defines the signs of $V$ and
$B$. Specifically, we can use this to interconvert the domain wall
configurations of sections I.A and I.B.  To keep track of the signs of
$V$ and $B$ in the left and right spaces we use a shorthand notation
${\cal H}[v_-,v_+;b_-,b_+]$ where $v_\pm$ and $b_\pm$ specify the
asymptotic signs of the potential and magnetic field strength. In this
notation, ${\cal H}[-,+;+,+]$ denotes a situation with $V<0$ on the
left and $V>0$ on the right, all immersed in a uniform positive field
$B>0$. We now introduce a local gauge transformation
\begin{equation}
{\cal S}(x) = \cos \alpha(x) \, \mathbb{I}  + i \sin \alpha(x) \, {\cal S}_1\:,
\label{local_gauge_transf}
\end{equation}
where $\alpha(-\infty) \rightarrow \pi/2$,
$\alpha(\infty) \rightarrow 0$ and $\alpha(0) = \pi/4$. This has the
effect of implementing a one-sided particle-hole transformation, where
the local gauge transformation evolves smoothly through the
interface. We retain the original problem for $x \gg 0$ but swap
particle and hole amplitudes for $x \ll 0$ to invert the signs of $V$
and $B$. This transformation is unitary and performs the mapping
\begin{eqnarray}
{\cal S} {\cal H}[-,+;+,+] {\cal S}^\dag = {\cal H}[+,+;-,+]\:,
\label{mapping}
\end{eqnarray}
thereby swapping the representation of a $pn$ junction in a uniform
field with a system with uniform doping in an antisymmetric $B$
field. In the doubled space the ground states can be identified
implying that the zero-mode structure is unchanged.  In
Fig.~\ref{fig7} we overlay the spectra calculated for the two
problems in the Nambu representation, illustrating this
correspondence.

\section{Pair Field in the interface}
\label{pair_field_section}
An interesting consequence of this mapping is the
structure of the spectrum inside the domain wall. The control
parameter $\alpha$ varies smoothly through the wall and has the
value $\alpha(0) = \pi/4$ exactly at its center. The
transformation in Eq.~(\ref{local_gauge_transf}) when
$\alpha = \pi/4$ is
\begin{eqnarray}
{\cal S}_d &=& ( \mathbb{I} + i \sigma_3 \otimes \Sigma_1)/\sqrt{2} \nonumber\\
 &=& \frac{1}{\sqrt{2}} \left(
                          \begin{array}{cccc}
                            1 & 0 & i & 0 \\
                            0 & 1 & 0 & -i \\
                            i & 0 & 1 & 0 \\
                            0 & -i & 0 & 1 \\
                          \end{array}
                        \right)\:.
\label{alpha_pi_over4}
\end{eqnarray}
As an example, for graphene we explicitly show
the complex structure of its sublattice off-diagonal terms:
\begin{eqnarray}
{\cal H}_{k} = \left(\begin{array}{cc}
      V & \gamma_k^* \\
      \gamma_k & V \\
     \end{array} \right)\:.
\end{eqnarray}
Inserting this in Eq.~(\ref{H_Nambu}), in the domain wall the
transformed Hamiltonian has the form
\begin{eqnarray}
\tilde {\cal H} &=& {\cal S}_d \cdot {\cal H}_N(k) \cdot {\cal S}_d^\dag \nonumber\\
  &=& \left(
        \begin{array}{cccc}
          0 & {\rm Re}[\gamma] & -iV & {\rm Im}[\gamma] \\
          {\rm Re}[\gamma] & 0 & {\rm Im}[\gamma] & iV \\
          iV & {\rm Im}[\gamma] & 0 & -{\rm Re}[\gamma] \\
          {\rm Im}[\gamma] & -iV & -{\rm Re}[\gamma] & 0 \\
        \end{array}
      \right)\:.
\end{eqnarray}
Thus, there is a pairing amplitude $\hat \Delta[V,\gamma_2]$ defined
by the local potential $V$ and the {\it imaginary part} of the hopping
amplitude $\gamma_2 = {\rm Im}[\gamma]$.  It is a matrix pairing
operator acting on the sublattice degrees of freedom.  It is symmetric
and thereby represents an effective p-wave pairing.

The appearance of an interfacial pair amplitude in the transformed
problem can be interpreted as follows. Suppose we introduced the local
gauge transformation in the absence of a domain wall. Then in the
particle language, electrons would pass through the putative interface
undeflected. But in the transformed language this means that an
incident electron is converted into a hole with unit probability. This
is an Andreev process that requires a pairing field in the Bogoliubov
Hamiltonian.

Note that a relation of this form indicates a formal mapping between
the problem of Klein tunneling~\cite{b5} and Andreev reflection at a boundary in
the Dirac theory. A closely related observation was made by Beenakker
and colleagues a few years ago~\cite{beenakker} who noticed that the
reflection amplitude for electrons incident on a symmetric graphene
$pn$ junction from the $n$-doped side has a precise analogy with a
problem where the particles were actually being Andreev reflected by a
contact with a superconductor. This work did not present the problem
in a Nambu basis and did not access the physics that we discuss below.

We now look at the symmetries of the original and transformed problem
in the domain wall. The original problem can be expressed,
using $\gamma_1 = {\rm Re}[\gamma]$, $\gamma_2 = {\rm Im}[\gamma]$,
and suppressing the $\otimes$, as
\begin{eqnarray}
{\cal H}_d = V \sigma_0\Sigma_3 + \gamma_1
  \sigma_1\Sigma_3 + \gamma_2 \sigma_2 \Sigma_0\:,
\label{H_domain}
\end{eqnarray}
while after the transformation we have
\begin{eqnarray}
\tilde {\cal H}_d = V  \sigma_3\Sigma_2 + \gamma_1 \sigma_1 \Sigma_3
+ \gamma_2 \sigma_1 \Sigma_1\:.
\end{eqnarray}
In both cases the ``$V$" term commutes with the $\gamma_1$ and
$\gamma_2$ terms while the $\gamma_1$ and $\gamma_2$ terms
anticommute.  For the $pn$ junction problem, $V=0$ in the interface.
For the antisymmetric $B$ problem $V$ is constant and can be absorbed
in the chemical potential. Then because of the anticommutation rules
for the $\gamma$ terms in Eq.~(\ref{H_domain}), zero-energy solutions
can only occur when $\gamma_1$ and $\gamma_2$ simultaneously vanish,
as they do precisely at the (projected) Dirac point. In the
transformed language this means that a zero-energy solution in the
Bogoliubov spectrum for the pseudo-superconductor will similarly
require that $\gamma_2 \rightarrow 0$ which reveals a momentum-space
linear node in the pairing field, consistent with its p-wave symmetry.

The Bogoliubov spectrum is invariant under global $U(1)$ gauge
transformations of its matrix-valued pair field $\hat \Delta$ in the
interface
\begin{eqnarray}
\hat \Delta'(\theta) = e^{-i \theta} \hat \Delta(0)\:.
\end{eqnarray}
This gauge degree of freedom can be identified with the continuous
family of possible transformations in Eq.~(\ref{S_theta}) that
interchange the particle and hole degrees of freedom in the Nambu
representation.  Thus any global $U(1)$ gauge transformation at the
interface ($x=0$) can be absorbed in a redefinition of the global
phase angle $\theta$ that defines the rotation that is used to switch
the particle and hole subspaces. Note that this choice is invisible in
the two bounding states but it does appear in the theory of the
interface.  However, a variation of the phase $\theta(y)$ along the
interface is a gauge choice and it is not associated with a physical
charge current along its tangent line.  To see this we observe that if
$\theta$ is promoted to a local (i.e. $y$-dependent) $U(1)$ degree of
freedom it also can be eliminated by a local $y$-dependent gauge
transformation back to a number conserving representation at the
expense of introducing a connection for operators that transport
electrons (holes) along the $y$ direction.  Here one can verify that
for generic Hamiltonians in the form $\Psi^\dag_i \Psi_j + \Psi^\dag_j
\Psi_i$ the generalization of Eq.~(\ref{alpha_pi_over4})
\begin{eqnarray}
{\cal S}_d(y) &=& ( \mathbb{I} + i \sigma_3 \otimes (\Sigma_1 \cos \theta(y) + \Sigma_2 \sin \theta(y)))/\sqrt{2} \nonumber\\
\end{eqnarray}
satisfies
\begin{eqnarray}
(\partial_y S_d^\dag) S_d + S_d^\dag \partial_y S_d = 0\:,
\end{eqnarray}
so that any variation $\theta(y)$ in the interfacial pair field
disappears completely and leaves no residual signature in the
number-conserving blocks of the back-transformed Hamiltonian. The
interfacial problem presents the more interesting case of a chiral
theory where the Hamiltonian instead has the structure
$-i \Psi^\dag \partial_y \Psi$. Here the connection introduces terms
in the Hamiltonian
\begin{eqnarray}
-i S_d^\dag \partial_y S_d = \left[ \Sigma_3 + (\Sigma_+ e^{-i\theta} + \Sigma_- e^{i \theta}) \sigma_3 \right] \partial_y \theta /2\:.
\end{eqnarray}
The first term commutes with the number conserving blocks and leads to
a momentum boost of the spectrum $\propto \partial_y \theta$. The
remaining terms anticommute with the number conserving blocks so their
effects appear only at higher order ${\cal O}(\partial_y \theta)^2$
and are unimportant for smoothly varying $\theta(y)$.  The momentum
boost has no effect on the boundary currents which depend on the
number of zero-energy intersections of the chiral branch of the
Hamiltonian. The boost can be interpreted as producing a valley
polarization within the occupied manifold, however lacking a sharp
definition of this polarization (the valleys are connected in the full
lattice theory), we expect that any variation $\theta(y)$ has no
effects on physically measurable quantities.

\section{Topological classification}

Particle-hole doubling of the domain wall problem promotes its
Hamiltonian to a Bogoliubov de Gennes (BdG) symmetry class. In the
original number-conserving representation, this system is gapped
breaking time-reversal symmetry ($B \neq 0$), particle-hole symmetry
and chiral symmetry ($\mu \neq 0$).  It is described by the
Altland-Zirnbauer~\cite{altland2005,schnyder2008,ryu2010,ezawa2013}
unitary symmetry class A which supports topologically nontrivial
ground states in two dimensions which are indexed by an integer-valued
invariant $\mathbb{Z}$. The count of the interface states in Sections
I.A and I.B
manifests the mismatch $\Delta \mathbb{Z} =2$ for the two
bounding gapped states across the domain wall.

The Nambu-doubled version of this problem explicitly restores
particle-hole symmetry $\Xi$ so that the extended problem can be
described by either Altland-Zirnbauer symmetry classes D or C,
depending on whether $\Xi^2$ equals $1$ (Class D) or $-1$ (Class C).  For
our application this distinction is important since in the former case
the ground state becomes topologically trivial and does not support
any symmetry protected modes at its edges or the interfaces. In the
latter case (C) the ground state has a $2 \mathbb{Z}$ topological
classification requiring an {\it even} number of interfacial/edge
modes in the spectrum. In either case an {\it odd} number of
topologically protected edge modes is excluded.  Combining the two
conditions
\begin{eqnarray}
\Xi^\dag \Xi = \mathbb{I} \nonumber\\
\{\Xi,\tilde {\cal H} \}= 0\:,
\end{eqnarray}
we have $\Xi = i \sigma_1 \Sigma_2$, so that $\Xi^2 = -1$ and the
extended problem is a member of class C.

We have verified this by an explicit calculation of the winding number
for the ground state. We carry this out in the number conserving
representation where the topological invariants from the particle and
hole sectors can be summed.  The Chern number ($\mathcal{C}$) for the
ground state can be calculated by choosing a chemical potential lying
in the first fundamental gap between Landau levels and summing up the
Chern numbers for all the individual bands below the chemical
potential. The Chern number of the $n$th Bloch band is
\begin{equation}
 \mathcal{C}_n=\frac{1}{2 \pi i} \int_{T^2} d^2k \left( \partial_{k_x} \mathcal{A}_{y}({\vec k})- \partial_{k_y} \mathcal{A}_{x}({\vec k})\right),
\end{equation}
where the Berry connection
\begin{equation}
\mathcal{A}_{\eta}({\vec k})=\langle n(k_x,k_y)| \partial_{k_\eta}| n (k_x,k_y)\rangle\:,
\end{equation}
for $\eta\in\{x,y\}$. Here $|n(k_x,k_y)\rangle$ represents the
normalized wavefunction of the $n$th Bloch band. We compute the Chern
number by discretizing the Brillouin zone (BZ) and summing up the
Berry curvature defined on each of the discretized
plaquettes\cite{fukui2005}. After discretization, the Chern number for
the $n$th band is
\begin{equation}
\tilde{\mathcal{C}}_n=\frac{1}{2\pi i}\sum_{k_x,k_y} F_n(k_x,k_y)\:,
\end{equation}
where $k_x, k_y$ are within the first BZ,
\begin{eqnarray}
 F_n(k_x,k_y)=&&\ln\big(U_{k_x}(k_x, k_y)U_{k_y}(k_x+\delta k_x, k_y)\nonumber \\
 &&U_{k_x}(k_x, k_y+\delta k_y)^{-1}U_{k_y}(k_x, k_y)^{-1}\big)\:,
\end{eqnarray}
and
\begin{eqnarray}
U_{k_x}(k_x, k_y)&=&\frac{\det\left( \Lambda^\dagger(k_x+\delta k_x,k_y)\Lambda(k_x,k_y)\right)}{|\det\left( \Lambda^\dagger(k_x+\delta k_x,k_y)\Lambda(k_x,k_y)\right)|} \nonumber \\
U_{k_y}(k_x, k_y)&=&\frac{\det\left(\Lambda^\dagger(k_x,k_y+\delta k_y)\Lambda(k_x,k_y)\right)}{|\det\left(\Lambda^\dagger(k_x,k_y+\delta k_y)\Lambda(k_x,k_y)\right)|}.
\end{eqnarray}
The column vectors of the matrix $\Lambda(k_x,k_y)$ are given by the
Bloch eigenstates $|n(k_x,k_y)\rangle$, where we include all the Bloch
bands below the chemical potential.  The Bloch eigenstates
$|n(k_x,k_y)\rangle$ are obtained by numerically diagonalizing the
tight-binding Hamiltonian for the graphene lattice with
nearest-neighbor hopping, where the phase of the hopping amplitude is
determined via the Peierls substitution.  Assuming that system is of
$n$-type and the magnetic field is uniform, adding up the Chern
numbers for the particle sector $(1)$ and the hole sector $(1)$, we
find that in the ground state $\mathcal{C}= 1 + 1 = 2$.  Similarly, we
find that $\mathcal{C}=-2$ for a $p$-type system in a uniform magnetic
field. This mismatch of Chern numbers suggests that there should be
four topologically protected edge states at the interface defining the
$pn$ junction. However, since we have artificially doubled the
spectrum by going to the Nambu basis, the genuine number of edge
states at the interface is two.

The following thought experiment provides an alternative route to
demonstrating the equivalence of the $[v_-,v_+,b_-,b_+] =[+,-,+,+]$
and $[+,+,+,-]$ interfaces.  The two situations are distinguished by
the substitution of $[v_+>0,b_+<0]$ by $[v_+<0,b_+>0]$ in the right
hand space. This replacement simultaneously negates the field
direction {\it and} the chemical potential and therefore it leaves the
Chern number of the gapped ground state unchanged. This substitution
can also be regarded as resulting from the gauge transformation in
Eq.~(\ref{mapping}) imposed on a uniform $(V,B)$ state and implemented
on a line {\it displaced from the physical interface}. Thus this
second wall is topologically trivial and does not support any
protected interfacial modes. Reducing the separation between the
original (physical) $[+,-,+,+]$ interface and the second (fictitious)
interface to zero generates the $[+,+,+,-]$ problem without changing
the pattern of domain wall currents.

\section{Variational Solution for the Interface}

Although the full energy spectrum of the graphene honeycomb lattice
can be calculated using a tight-binding approximation, additional
insight can be obtained from the continuum approximation. In this case
a Taylor expansion centered at two non-equivalent Dirac points
$\mathbf{K}$ and $\mathbf{K'}$ produces the Dirac-like Hamiltonians
$H_{K}=v_{F} \left(\sigma_x \tau_z \hat{p}_x + \sigma_y \hat{p}_y
\right)$
where $\tau_z$ acts on the $K$ and $K'$ valley indices and $v_{F}$ is
the Fermi velocity. Together these Hamiltonians can be combined to
reproduce a $4\times 4$ two-valley Hamiltonian in terms of the Dirac
gamma matrices\cite{d4}.  Due to the similarities of the two
Hamiltonians, here we will only consider the $\mathbf{K}$ point where
the Hamiltonian with an external potential $V(x)$ becomes~\cite{b2,b3,b4}
\begin{equation}
\hat{H}=v_F (\vec{\sigma}\cdot\hat{p})+V(x) \mathbb{I}\:,
\label{hamiltonian} 	
\end{equation}
where $\vec \sigma$ are Pauli matrices in the sublattice
basis, $\hat{p}= (\hat{p}_{x},\hat{p}_{y})$ is the momentum operator,
$\mathbb{I}$ is the identity matrix, and $V(x)$ is an external or bias
potential associated with an applied electric field.  Below we
consider the two cases of antisymmetric electric and symmetric
magnetic fields and vice versa. In either case the vector potential is
written in Landau gauge with $\vec{A} = A_y(x) \hat y$ and the
Dirac Hamiltonian associated with the $K$ valley and normalized for
$v_F$ has the form
\begin{eqnarray}
H= \left(\begin{array}{cc}
      V(x) & -i \partial_x - (\partial_y -iA_y) \\
      -i\partial_x+ (\partial_y-iA_y) & V(x) \\
     \end{array} \right).
\label{valley_hamiltonian}
\end{eqnarray}
The solution for the Dirac equation $ H\Psi=\epsilon \Psi$ can be
written in the Bloch form
\begin{equation}
	\psi=
	e^{ik_{y}y}
	\left(\begin{array}{ccc}
		a(x)\\
		b(x)
	\end{array}\right)
	\label{psi}
\end{equation}
indexed by the conserved wave vector $k_y$. After this substitution
the Dirac equation is rewritten as a pair of coupled second-order
equations:
\begin{eqnarray}
\left( \left[ -\partial_x^2 +(k_y - A_y)^2 \right] \hat{\mathbb{I}}  - \partial_x A_y \sigma_z \right) \psi \nonumber\\
= \left( (\epsilon-V)^2 \hat{\mathbb{I}}  + i \partial_x V \sigma_x \right) \psi
\end{eqnarray}
which defines a complex matrix-valued potential
\begin{eqnarray}
\mathbb{U} = (k_y - A_y)^2  \hat{\mathbb{I}}  - B(x) \sigma_z - i \partial_x V \sigma_x\:.
\label{matrix_potential}
\end{eqnarray}

\begin{figure}
  \includegraphics[angle=0,width=\columnwidth]{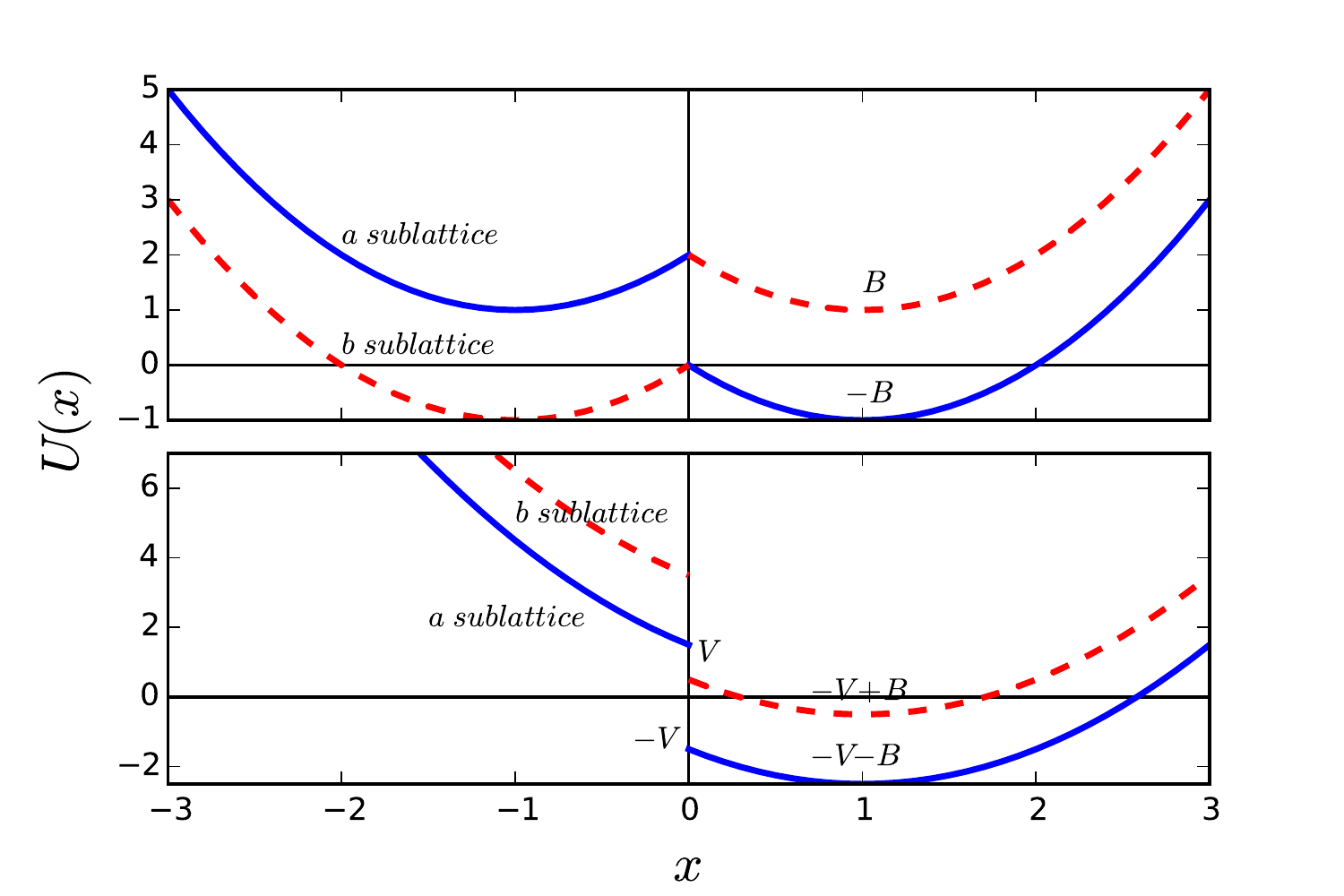}
  \caption{\label{fig8} Upper panel: potential of the interface
    created by an antisymmetric magnetic field
    $B(x) = B\: {\rm sign}(x)$. For both sublattices we have a
    pseudo-Zeeman shifted double parabolic potential.  Lower panel:
    potential of the interface created by an antisymmetric electric
    bias field $V(x) = - V {\rm sign}(x)$. It consists of one
    parabola, which has a jump at the interface related to the value
    of the electric field in the $pn$ junction, e.g. as
    $E=\partial_x V(x)$.}
\end{figure}

In the case of an antisymmetric $B$ field with uniform doping,
$\partial_x V=0$ and these equations are decoupled in the sublattice
basis so they can be solved separately. For both sublattices we have a
particle described by a massive Schr\"odinger-like equation in a
double parabolic potential. In the case when
$B(x)= B\: {\rm sign} (x)$ the potentials are shown in the upper panel
of Fig.~\ref{fig8}. The minima in these potentials occur
at the values $x = \pm k_y/B$, offset in energy by the pseudo-Zeeman
$\sigma_z$ term in Eq.~(\ref{matrix_potential}). The $a$ and $b$
sublattice potentials are exchanged by the reflection
$x \rightarrow -x$.

Equation~(\ref{valley_hamiltonian}) reveals that these two amplitudes
are not independent, but are coupled through the
$-i \sigma_x \partial_x$ term in the linearized Hamiltonian.  The form
of the effective potential $\mathbb{U}$ suggests a useful variational
basis for studying the states bound to the domain wall,
\begin{eqnarray}
a(x) = C e^{-(x-x_a)^2/\sigma^2} = C e^{-(x-d)^2/\sigma^2} \nonumber\\
b(x) = C e^{-(x-x_b)^2/\sigma^2} = C e^{-(x+d)^2/\sigma^2},
\end{eqnarray}
which are two overlapping Gaussians each of width $\sigma$ and
separated by $2d$; $C$ is a normalization factor. In this variational
space the off-diagonal coupling describes a tunneling between states
on opposite sides of the interface and with opposite sublattice
polarizations. Evaluating the matrix element one finds
\begin{eqnarray}
\langle a | {\cal H} | b \rangle &=& - \langle b | {\cal H} | a \rangle \nonumber\\
      &=&-i \left( \frac{2d}{\sigma^2} - k_y + B \langle |x|
          \rangle\right)  
e^{-2d^2/\sigma^2}  \nonumber\\
 &\approx& -i \left( \frac{2d}{\sigma^2} \right)  e^{-2d^2/\sigma^2}  
\label{h_ab}
\end{eqnarray}
which is optimized for $d = \sigma/2$ in a gauge where a minimum in
the effective potential occurs at $x=0$ when $k_y=0$.
This demonstrates that the localized variational states are most
effectively tunnel-coupled at a finite range from the domain wall. The
charge density in the lowest energy state is therefore sublattice
polarized on opposite sides of the interface as seen in the charge
density plotted in Fig.~\ref{fig2}.  The tunneling
Hamiltonian in this variational basis is proportional to
$\sigma_y$. The energy-optimized state is a coherent combination of
these trial basis states forming an eigenfunction of the velocity
operator $v_F \sigma_y$ with negative eigenvalue: this is a domain
wall state propagating along the $-\hat y$ direction. The same
physics occurs in the $K'$ valley. In either case the variational
solution describes a situation where an incident electron bound in a
cyclotron orbit is transmitted through the domain wall where the field
is reversed producing a counter circulation of the orbit and
liberating the average motion along the tangent line of the wall.

In the case of a homogeneous magnetic field and antisymmetric doping,
when $V(x)=-V(-x)$, the $a$ and $b$ sublattice solutions are coupled
by the $-i \partial_x V \: \sigma_x$ term in the potential
Eq.~(\ref{matrix_potential}). For the specific example considered
here, when $V(x)= V {\rm sign}(x)$ this coupling appears as boundary
conditions at the interface,
 \begin{eqnarray}\label{bca-2}
-\partial_x a(0+) +\partial_x a(0-)  = iv b(0) \nonumber\\
-\partial_x b(0+) +\partial_x b(0-)  = iv a(0)\:,
\end{eqnarray}
where $v$ is the jump of the electrostatic potential at the
interface. The potentials in shown in the lower panel of
Fig.~\ref{fig8} again suggest a variational basis.  Here
the dominant low-energy degree of freedom is $a$-sublattice polarized
in the lowest Landau level on the right side (low potential side) of
the interface. It is coupled to an evanescent mode on the right which
is $b$-sublattice polarized penetrating a barrier from the
pseudo-Zeeman field, and to evanescent modes on the left penetrating
the electrostatic barrier on both sublattices.  This physics is
captured in the variational basis
\begin{eqnarray}
a(x) &=& C_a e^{-(x-d)^2/\sigma^2}; \,\, (x>0) \nonumber\\
     &=& C_a' e^{x/\ell}; \,\, (x<0) \nonumber\\
b(x) &=& C_b e^{-x/\ell_c}; \,\, (x>0) \nonumber\\
     &=& C_b e^{x/\ell}; \,\,\, (x<0)
\label{var_basis}
\end{eqnarray}
where $\ell_c$ and $\ell$ are decay lengths for penetrating the 
pseudo-Zeeman barrier and the electrostatic barrier, respectively. The
matching condition can be written in matrix form
\begin{eqnarray}
\left(
  \begin{array}{cc}
    1/\ell - 2d/\sigma^2 & -iv \\
    -iv & 1/\ell + 1/\ell_c \\
  \end{array}
\right) \left(
          \begin{array}{c}
            a(0) \\
            b(0) \\
          \end{array}
        \right)
= 0\:.
\label{matching}
\end{eqnarray}
A nontrivial solution occurs when the determinant of the matching
matrix is zero, giving
\begin{eqnarray}
d = \frac{\sigma^2}{2 \ell} \left( 1 + v'^2 \ell \tilde{\ell} \right)
\end{eqnarray}
where $\tilde \ell = \ell \ell_c/(\ell + \ell_c)$.  The interfacial
mode is a vector in the nullspace of Eq.~(\ref{matching})
\begin{eqnarray}
\psi(0) = C \left(
            \begin{array}{c}
              1 \\
              iv \tilde{\ell} \\
            \end{array}
          \right)\:.
\end{eqnarray}
Here, $v < 0$ so the group velocity is directed along the $-\hat y$
direction. This velocity is evidently proportional to $v$, consistent
with its identification with the classical drift velocity $E/B$
expected in the classical theory of an edge/interface state.
 
\begin{figure}
  \includegraphics[angle=0,width=\columnwidth]{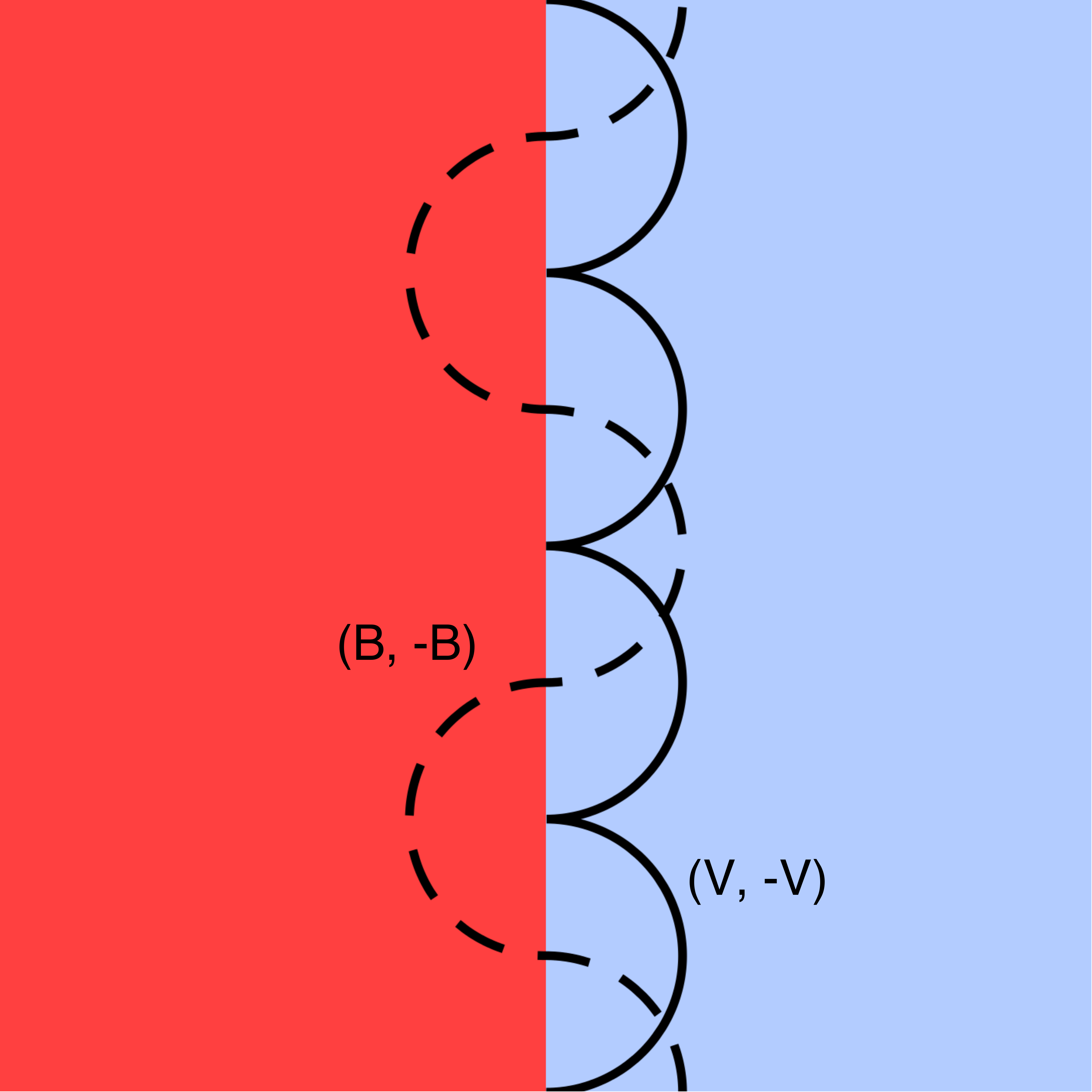}
  \caption{\label{fig9}
The two different types of domain walls
support two different kinds of semiclassical trajectories. Solid line:
skipping orbits typical for the uniform magnetic
field case at a $(V,-V)$ domain wall. Dashed lines: snake orbits
typical for a $(B,-B)$ domain wall.}
\end{figure}

Note also the sublattice asymmetry in this boundary solution: the
interfacial mode is not in a velocity eigenstate as was found in the
antisymmetric $B$ problem and has its dominant amplitude on the
sublattice found in the zero-energy mode (lowest Landau level) in each
valley.  This is clearly seen in the charge densities calculated for a
version of this interface plotted in Fig.~\ref{fig5}. The evanescent
form of the trial functions in Eq.~(\ref{var_basis}) reflects the
different physics operative in the electrostatic barrier. Here an
electron bound in a cyclotron orbit is backscattered by an electric
field at the interface, weakly penetrating both an electrostatic
barrier and a Zeeman barrier due to the sublattice polarization. This
behavior is illustrated in Fig.~\ref{fig9}.

\section{Discussion}

Graphene interfaces that support electrostatic barriers (at a $pn$
junction) or field reversal (at a magnetic domain wall) can host
confined snake state solutions. The dynamics responsible for these
solutions is clearly different in these two situations. For the
magnetic domain wall cyclotron orbits of opposite circulation are
matched on a boundary to produce an unbound propagating
excitation. For the electrostatic barrier a transverse electric field
is introduced that reflects an incident mobile carrier to produce a
skipping orbit drifting at velocity $E/B$.  Despite these differences
the pattern of boundary currents is the same demonstrating the common
topological character of both interfaces. We found that these two
problems can be mapped onto each other using a particle-hole doubled
representation. In this extended basis the two problems are
interconverted by a local gauge transformation using the particle and
hole degrees of freedom in the Nambu basis. An interesting and
unavoidable consequence of this mapping is that a number-conserving
version of one problem is the image of a Bogoliubov de Gennes problem
for fermions with an interfacial pair field in the other. This
relation offers the interesting possibility of generalizing coherent
quantum transport phenomena such as Andreev reflection or Veselago
lensing, to a new family of graphene-derived architectures.

Our results imply two ``no-go" theorems for this system. First, the
gauge-transformed representation of the interfacial problem describes
a one-dimensional particle/hole gas coupled by a pairing field with
p-wave symmetry. In principle, this class of models can support
Majorana excitations in an appropriate parameter
regime~\cite{kitaev01,tiwari2013}.  However, the prospects for realizing such
excitations in this setup are remote. We find that the doubled problem
has the discrete symmetries of Altland-Zirnbauer symmetry class C so
that its gapped ground state is indexed by an {\it even}
integer-valued ($2\mathbb{Z}$) index.  Here topological domain wall
solutions must appear in pairs, and the possibility of having an
unpaired Majorana excitation appearing on the boundary, or at its ends
are excluded.  This is understandable, since the appearance of such an
excitation would not admit an interpretation in the original (number
conserving) representation of the same problem. Second, it is
intriguing that under the gauge transformation the interfacial problem
exhibits an apparent broken $U(1)$ gauge symmetry due to the
pseudo-pairing field. However, this seems not to be associated with
any measurable collective effects in the interface: spatial variations
of the phase of the order parameter are the images of momentum shifts
of the spectrum in the original number-conserving representation of
the problem, which is simply a gauge choice. Again, any nontrivial
property arising from the pairing field would require a dual
interpretation in the number-conserving representation of the same
problem. It remains an open question as to whether one might further
break the symmetries of the original problem to identify measurable
consequences of these symmetries in its gauge transformed image.

Our results highlight several directions for exploiting the structure
of the domain wall solutions.  It is possible that the sublattice and
valley asymmetries found for two problems could be exploited to valley
filter ballistic transport in patterned graphene. The two-channel
interfacial solutions may also provide a venue for important
one-dimensional interaction effects that are accessible in nonlocal
transport measurements between reservoirs bridged by a domain
wall. Finally, we also note that while graphene provides a natural
starting point for developing this formulation, the conclusions appear
to be generally valid for ballistic transport in domain walls
separating topologically gapped ground states in a wider class of
semiconductor nanostructures.

\begin{acknowledgments}
  We would like to acknowledge fruitful discussions with P. Makk,
  P. Rickhaus, C. Sch\"onenberger, and C.L. Kane.  The work of
  C.B. and R.P.T. was financially supported by the Swiss SNF and the
  NCCR Quantum Science and Technology.  E.J.M. is supported by DOE
  grant DE-FG02-ER45118 and a grant from the Leverhulme Foundation at
  Loughborough University while this work was carried out.
\end{acknowledgments}

\end{document}